# PET Tracer Separation Using Conditional Diffusion Transformer with Multi-latent Space Learning

Bin Huang, Feihong Xu, Xinchong Shi, Shan Huang, Binxuan Li, Fei Li, Qiegen Liu, *Senior Member, IEEE*

*Abstract*—In clinical practice, single-radiotracer positron emission tomography (PET) is commonly used for imaging. Although multi-tracer PET imaging can provide supplementary information of radiotracers that are sensitive to physiological function changes, enabling a more comprehensive characterization of physiological and pathological states, the gamma-photon pairs generated by positron annihilation reactions of different tracers in PET imaging have the same energy, making it difficult to distinguish the tracer signals. In this study, a multi-latent space guided texture conditional diffusion transformer model (MS-CDT) is proposed for PET tracer separation. To the best of our knowledge, this is the first attempt to use texture condition and multi-latent space for tracer separation in PET imaging. The proposed model integrates diffusion and transformer architectures into a unified optimization framework, with the novel addition of texture masks as conditional inputs to enhance image details. By leveraging multi-latent space prior derived from different tracers, the model captures multi-level feature representations, aiming to balance computational efficiency and detail preservation. The texture masks, serving as conditional guidance, help the model focus on salient structural patterns, thereby improving the extraction and utilization of fine-grained image textures. When combined with the diffusion transformer backbone, this conditioning mechanism contributes to more accurate and robust tracer separation. To evaluate its effectiveness, the proposed MS-CDT is compared with several advanced methods on two types of 3D PET datasets: brain and chest scans. Experimental results indicate that MS-CDT achieved competitive performance in terms of image quality and preservation of clinically relevant information. Code is available at: https://github.com/yqx7150/MS-CDT.

This work was supported by National Natural Science Foundation of China (Grant number 62122033, 62201193), and in part by Nanchang University Interdisciplinary Innovation Fund Project (Grant number PYJX20230002). (B. Huang and F. Xu are co-first authors.) (Corresponding authors: F. Li and Q. Liu.)

This work involved human subjects or animals in its research. Approval of all ethical and experimental procedures and protocols was granted by the Ethics Committee of Clinical Drugs, Equipment and New Medical Technologies of the First Affiliated Hospital of Sun Yat-sen University.

B. Huang and F. Xu are with School of Mathematics and Computer Sciences, Nanchang University, Nanchang, China ({huangbin, 406100230043}@email.ncu.edu.cn).

X. Shi is with First Affiliated Hospital of Sun Yat-Sen University, Department of Nuclear, China (shixch@mail.sysu.edu.cn).

B. Li is with Institute of Artificial Intelligence, Hefei Comprehensive National Science Center, Hefei, China (libingxuan@iai.ustc.edu.cn).

F. Li and S. Huang are with Department of Nuclear Medicine, the Second Affiliated Hospital of Anhui Medical University, China (lifei007@139.com, huangs888vip@163.com).

Q. Liu is with School of Information Engineering, Nanchang University, Nanchang, China (liuqiegen@ncu.edu.cn).

## I. INTRODUCTION

Positron Emission Tomography (PET) is a medical imaging technique based on nuclear physics and molecular physics. It enables the observation of cellular metabolic activities at the molecular level and provides valid evidence for the detection and prevention of certain heart diseases, neurological disorders, and tumors. Compared with other imaging techniques, on one hand, there are various PET radioactive tracer drugs, and different tracers can characterize the abnormal uptake of different diseased cells, such as characterizing and quantifying glucose [1-3], blood flow [4-6], hypoxia [4, 7, 8], cell proliferation [9, 10], amino acid synthesis [11-13, gene expression [14, 15], and other physiological processes. On the other hand, PET imaging has high sensitivity and high specificity, especially in lesion detection, so it can be applied in many specific medical scans and tests, such as detecting small bone injuries [16], cancer bone metastases [17], brain tumors [18], and Parkinson's disease [19], etc. PET has become a valuable tool for clinical disease diagnosis and treatment [20][21]. This imaging technology enables the visualization and quantification of the body's metabolic processes, providing crucial information for accurate medical decision-making [22].

Traditional PET imaging relies on single-tracer PET imaging to provide detailed information about the physiological state of diseases. In clinical practice, each tracer can only represent the information within one type of cell, which may potentially lead to false negative or false positive diagnoses. In comparison, the application of multi-tracer imaging allows the simultaneous observation of multiple aspects of biological metabolism. Different tracers will automatically record information in a timely manner, and the multi-tracer PET imaging technology can provide complementary information [23] from radioactive tracers sensitive to different physiological function changes, enabling a more complete characterization of the disease state, thus improving the accuracy of disease diagnosis and reducing the risk of misdiagnosis. Moreover, multi-tracer imaging can be further divided into dynamic and static imaging. Dynamic PET parametric imaging with higher quantitative accuracy is more difficult to be commonly used in clinical practice due to issues such as longer scanning time, the need for multiple scans within a period of time, and the requirement of invasive sampling for some blood input functions. For example, Gao *et al.* conducted a synchronous dynamic tracer reconstruction of tissue activity maps guided by tracer kinetics. The tracer reconstruction problem was formulated in the state-space representation form [24]. Zeng *et al.* proposed a multi-task CNN, which is a three-dimensional neural network based on the multi-task learning framework. A shared encoder extracts features from the tracer dynamic sinograms, followed by two distinct parallel decoders that reconstruct the single tracer dynamic images of the two tracers, respectively [25]. In contrast, static PET imaging only involves a single scan at a specific time point, which reduces radiation exposure and discomfort for patients and has lower requirements for equipment. Hence, static multi-tracer imaging holds great promise.

Traditional approaches for separating tracer PET data, such as pixel-wise and region-based methods, have proven useful in several applications. However, these techniques often struggle with noise and poor spatial resolution, limiting their accuracy and reliability in complex clinical scenarios. Recent advancements in deep learning methods have shown great potential in addressing these challenges. CNNs, have demonstrated significant improvements in separating tracer PET images, enabling more precise delineation of metabolic activity and tracer distribution. For instance, Liu *et al*. [26] proposed a deep learning framework using a CNN to reconstruction tracer PET images by directly learning the underlying features of each tracer, thus reducing the influence of noise and improving image quality. The network was trained on paired datasets, allowing it to capture the distinct characteristics of each tracer while preserving spatial details. Similarly, Zeng *et al*. [25] introduced a novel multi-task learning-based method to separate tracer PET images, incorporating a shared encoder with separate decoders for each tracer. This architecture enabled the model to learn common features between the tracers while simultaneously focusing on tracer-specific patterns. Their approach demonstrated enhanced performance in terms of both image clarity and quantification accuracy when compared to traditional methods.

With the development of deep learning, both diffusion models and Transformer models have shown remarkable promise in the fields of tracer PET image reconstruction and denoising. They have demonstrated the effectiveness in addressing these challenges by improving image quality and preserving essential diagnostic details. Diffusion model is a generative model that learns the underlying data distribution by simulating a diffusion process. Recently, it has received attention for its ability to denoise and enhance PET images. Gong *et al*. [25] demonstrated the potential of using a denoising diffusion probabilistic model to reduce noise in PET images, indicating that better performance can be achieved by using MR priors as network inputs while embedding PET images as data consistency constraints. Similarly, Jiang *et al*. [26] proposed an unsupervised PET enhancement method based on a latent diffusion model, which effectively enhanced PET images with latent information. Besides, Transformer models are widely recognized for their ability to capture long-term dependencies and process large-scale data effectively. In the context of medical imaging, Transformer models have been utilized to improve tracer PET image reconstruction and denoising by leveraging their powerful feature extraction and representation capabilities. Zhang *et al*. [27] introduced the Spatial Adaptive and Transformer Fusion Network (STFNet) for blind denoising of low-count PET images. To the best of our knowledge, there is currently no relevant research on tracer PET separation.

Given the complementary advantages of diffusion models and Transformer models, some studies have already applied diffusion Transformer models to the PET field. Huang *et al*. proposed a diffusion Transformer model guided by a joint compact prior to improve the reconstruction quality of low-dose PET imaging [29]. Meanwhile, they also proposed integrating a mask mechanism from 2D to 3D into both the sinogram domain and the latent space for PET reconstruction [30]. In order to better combine with the dual-tracer problem, we introduce multiple latent space, with each latent space corresponding to one tracer. On this basis, we use the texture mask as a condition to further enhance the separation details.

In this study, we introduce a multi-latent space guided texture conditional diffusion transformer model (MS-CDT) for dual-tracer PET image separation. The model incorporates the texture mask condition (TMC) and the multi-latent space prior (MSP) corresponding to tracer PET. Through the texture mask conversion block (TMCB), the tracer PET is transformed into texture masks as conditional information. Then, through the latent prior extraction block (LPEB), the dual-tracer PET and single-tracer PET are rearranged and stitched at the pixel level. Features are extracted via a convolutional network and an encoder, and compressed into a low-dimensional space, significantly reducing the computational load. Moreover, the model combines the powerful distribution-mapping ability of the diffusion model and the capability of the transformer model to capture long-range dependencies. The diffusion model only needs to predict the MSP instead of entire dual-tracer data. The model is adaptable to the number of tracers, allowing the latent space structure to scale accordingly. The present study focuses on the dual-tracer PET separation problem. The TMC and MSP are used as guidance during the training process of the transformer model, enabling efficient feature disentanglement and improving the separation accuracy.

The theoretical and practical contributions of this work are summarized as follows:
● TMC learning is introduced to alleviate the over smoothing problem commonly encountered in deep learning method. The proposed texture mask conversion module captures salient structural patterns and serves as conditional guidance within the diffusion transformer model. By incorporating these texture masks as structural conditions, the model is encouraged to preserve fine-grained details in the separation output, resulting in improved visual quality and structural fidelity of the separated tracer images.
● MSP representation is proposed for independent feature disentanglement, where each latent space encodes the characteristics of a specific tracer. This design enables the model to isolate and preserve tracer-specific semantic and structural information. By predicting and utilizing MSP rather than directly reconstructing the entire PET image, the model ensures higher separation fidelity while maintaining computational efficiency.

 The rest of the manuscript is organized as follows. Section II presents the relevant background of the diffusion model and the transformer model guided by the texture mask. The detailed steps and algorithms of these two methods are described in Section III. Section IV provides the experimental results and specifications regarding the implementation and experiments. Finally, conclusions are drawn in Section V.

## II. PRELIMINARY

### A. Transformer Model

Transformer is a deep learning architecture used for natural language processing and other sequence-to-sequence tasks [31]. It was first proposed in 2017 by Vaswani *et al*. Nowadays, it has been applied to various visual tasks, including image recognition, segmentation, and object detection. The Transformer model has strong parallel computing capabilities, excellent long-distance dependency capturing ability, and high versatility and extensibility. However, it consumes a large amount of computing resources and lacks an inherent perception of the input sequence order. The Transformer model has a relatively strong ability to model global dependencies. When dealing with complex spatial information, it can better capture the relationships between tracers and single tracers and conduct effective image separation. The components of the Transformer are as follows:

*Self-Attention Mechanism:* The self-attention mechanism enables the model to consider all positions in the input sequence simultaneously. It allows the model to assign different attention weights according to different parts of the input sequence, thereby better capturing semantic relationships.

*Multi-Head Attention Mechanism:* The self-attention mechanism in the Transformer is extended to multiple attention heads, allowing the model to process different information latent spaces in parallel. Each head can learn different attention weights to better capture different types of relationships.

*Feed Forward Neural Network:* After the multi-head attention mechanism, a fully connected feed-forward neural network is connected. It independently performs nonlinear transformations on the vectors at each position to further extract and integrate features, enhancing the model's ability to represent input information.

*Layer Normalization and Residual Connection:* After the multi-head attention mechanism and the feed-forward neural network, layer normalization operations are added. These operations normalize the input of each layer, which helps stabilize the training process and accelerate convergence. Meanwhile, there are also residual connections, which means directly adding the input of each layer to the output after processing. This helps solve problems such as vanishing gradients, enabling information to be transmitted more smoothly within the network and allowing the model to learn deep features more effectively.

### B. Diffusion Models

Diffusion models have become the forefront in density estimation [32] and sample quality enhancement [33]. The inspiration for diffusion models comes from non-equilibrium thermodynamics. Theoretically, Markov chains are first defined for diffusion steps to slowly add random noise to the data, and then the reverse diffusion process is learned to construct the desired data samples from the noise. These models utilize parameterized Markov chains to optimize the lower variational bound of the likelihood function, enabling them to generate a target distribution that is more accurate compared to other generative models.

The forward process of diffusion models, which operates on an input image $x_0$, gradually transforming it into Gaussian noise $x_t \sim \mathcal{N}(0, I)$ through $t$ iterations. Each iteration of this process is described as follows:

$$q(x_t|x_{t-1}) = \mathcal{N}(x_t; \sqrt{1-\beta_t}x_{t-1}, \beta_t I) \quad (1)$$

where $x_t$ denotes the noised image at time-step $t$, $\beta_t$ represents the predefined scale factor, and $\mathcal{N}$ represents the Gaussian distribution. During the reverse process, diffusion models sample a Gaussian random noise map $x_t$, then progressively denoise $x_t$ until it achieves a high-quality output $x_0$:

$$p(x_{t-1}|x_t, x_0) = \mathcal{N}(x_{t-1}; \mu_t(x_t, x_0), \sigma_t^2 I) \quad (2)$$

To train a denoising network $\epsilon_\theta(x_t, t)$, given a clean image $x_0$, diffusion models randomly sample a time step $t$ and a noise $\epsilon \sim \mathcal{N}(0, I)$ to generate noisy images $x_t$ according to Eq. (2).

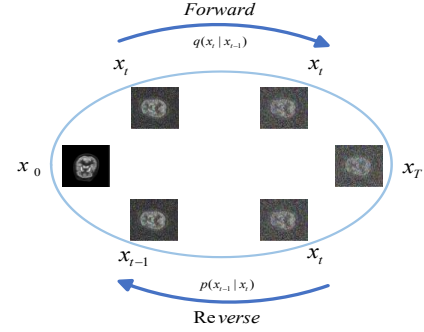

**Fig. 1.** Forward and reverse processes of diffusion models.

## III. PROPOSED METHOD

Both the diffusion model and the transformer model have their unique advantages: the diffusion model excels at distribution learning and generating fine-grained representations through iterative denoising, while the transformer model captures long-range dependencies via multi-head attention and supports efficient parallel processing. Based on these strengths, a diffusion transformer model (DTM) [29] is proposed. To better adapt DTM to PET tracer separation, we introduce a TMCB and a LPEB, and seamlessly incorporate them into the DTM backbone. The TMCB leverages texture-guided conditioning to highlight salient structural patterns, mitigating over smoothing and enhancing fine-detail preservation. The LPEB enables independent tracer-specific feature disentanglement by extracting and utilizing low-dimensional latent priors, thereby improving separation accuracy and computational efficiency.

This study is approved by the institutional review board of the Beijing Friendship Hospital, Capital Medical University, Beijing, China. The approval number is 2022-P2-314-01 and approval date is Oct. 8, 2022.

### C. MS-CDT

In summary, the present MS-CDT consists of three parts: the TMCB, LPEB, and the diffusion transformer model. As shown in Fig. 2, a TMCB is proposed to transform the dual-tracer PET into texture conditions. Secondly, through a LPEB, the dual-tracer PET and the single-tracer PET are converted into latent priors respectively. Subsequently, the MSP and TMC are compressed into a low-dimensional image space to reduce the computational complexity. The MSP and TMC can achieve independent feature disentanglement, enabling the model to more effectively decouple and extract the unique physical features of each PET tracer, thus improving the tracer separation accuracy. LPEB is mainly composed of stacked downsampling, residual blocks, linear layers, and pooling layers.

In the diffusion stage, the diffusion model learns MSP of the two tracers. Instead of predicting the prior information of the entire data, the diffusion process predicts the denoised dual-latent prior and the dual-texture condition through conditional diffusion. That is, the dual-latent space priors of the two tracers' PET obtained from the MSP extraction block are fed into the diffusion model for prediction. Then, the priors are input into the Transformer module as affine transformation parameters to reconstruct the separated two tracers' PET.

In the Transformer stage, guided by the texture conditions and priors, the Transformer directly estimates the image features by utilizing the two latent priors learned by the diffusion model, thus generating stable and realistic outputs. Multiple Transformer blocks are combined in the form of a U-net, and each block is composed of a multi-head attention mechanism and a feed-forward neural network. The Transformer module captures long-range pixel interactions. Among them, the multi-head transfer attention module aggregates local and non-local pixel interactions, demonstrating its ability to perform feature interactions across channels. The gated feed-forward network suppresses less informative features and only allows useful information to pass further through the network hierarchy. Finally, guided by the texture conditions and priors, the Transformer utilizes two priors to separate two tracer PETs from the dual-tracer PET.

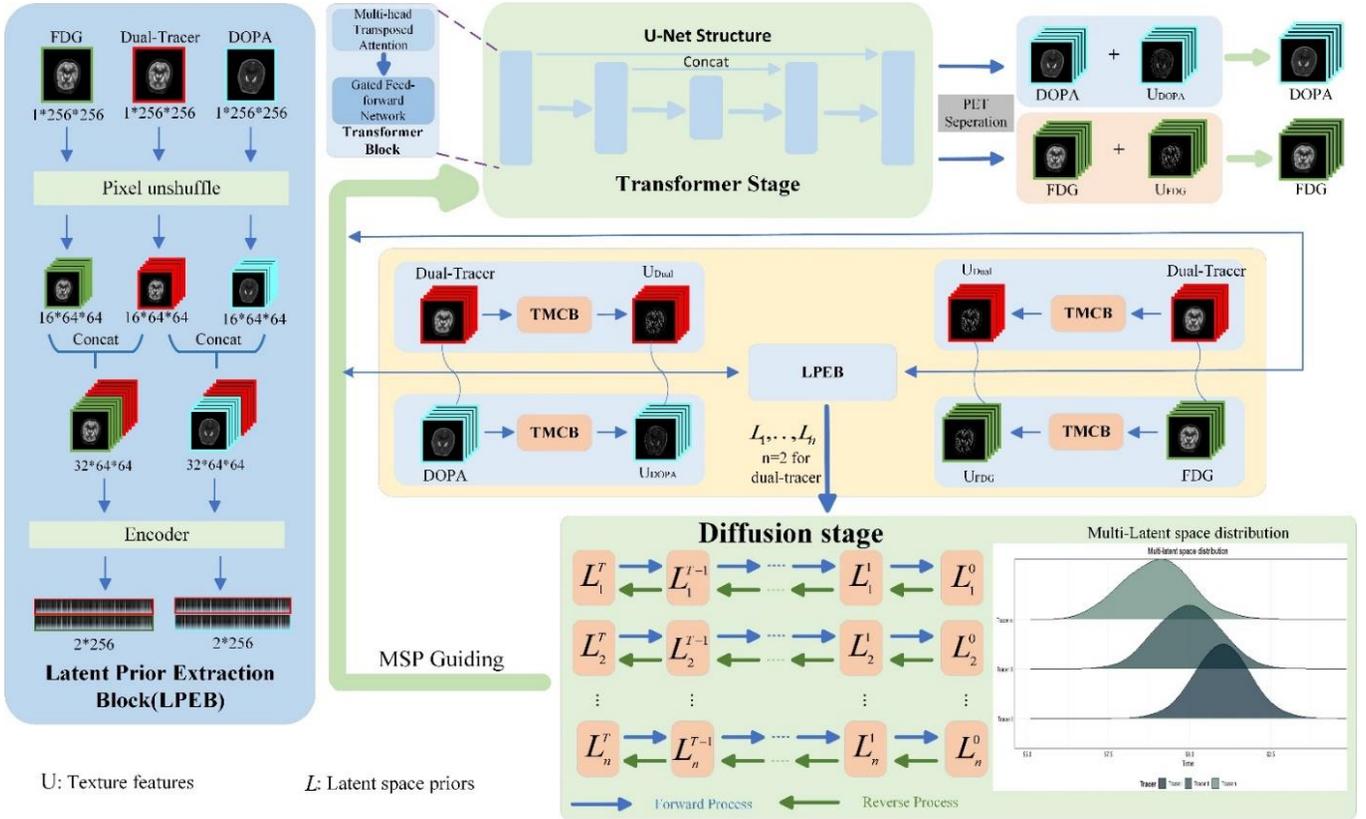

**Fig. 2.** Pipeline of the MS-CDT network architecture. MS-CDT is mainly composed of a TMCB, an LPEB, a diffusion stage, and a transformer stage. The tracer PET obtains texture features through the TMCB. The dual-tracer PET and the two single-tracer PETs are respectively combined into an MSP. The texture conditions and the MSP enter the diffusion stage, the texture conditions and the MSP guide the transformer stage to utilize the predicted priors for separation.

### D. TMCB

Texture is an important visual feature in images, which reflects the spatial distribution pattern of gray levels or colors. Texture is characterized by features such as repetitiveness, regularity, and directionality. TMCB utilizes the local binary pattern (LBP) operator to extract local texture information, which is robust to both rotation and gray-scale variations. This characteristic is especially beneficial for tracer PET signals, where anatomical texture may vary subtly across spatial regions. Specifically, the LBP operator is defined over a local 3×3 window. Taking the pixel at the center of the window as a threshold reference, each of the eight neighboring pixels is compared against it. If the gray value of a neighboring pixel is greater than or equal to that of the center pixel, it is assigned a value of 1; otherwise, it is assigned a value of 0. This comparison yields an 8-bit binary pattern that characterizes the local texture structure within the window. The LBP value of the pixel at the center of the window is computed as:

$$LBP(x_c, y_c) = \sum_{p=1}^{8} s(I(p) - I(c)) * 2^p \quad (3)$$

$$s(x) = \begin{cases} 1, & x \geq 0 \\ 0, & x < 0 \end{cases} \quad (4)$$

The TMC is then constructed by applying the LBP operator across the entire tracer PET image $I$, such that:

$$TMC = \begin{cases} 1, & if\ LBP(x,y) \geq 0 \\ 0, & if\ LBP(x,y) < 0 \end{cases} \quad (5)$$

These texture masks serve as condition vectors that guide both the diffusion and transformer modules, providing fine-grained structural priors. The texture image is computed as:

$$U = I \odot TMC \quad (6)$$

Subsequently, $U$ is multiplied with the tracer signal to preserve the texture information within the masked regions. This refined texture information is then fed into the diffusion transformer model for condition learning. During inference, the final separated image is obtained via weighted fusion of the output image $\hat{I}$ and the masked texture image $\hat{U}$, as follows:

$$\hat{I}_{final} = \alpha \cdot \hat{I} + (1-\alpha) \cdot \hat{U} \quad (7)$$

This fusion mechanism effectively compensates for potential texture loss introduced by deep models, thereby alleviating the over-smoothing problem and enhancing the preservation of fine structural details.

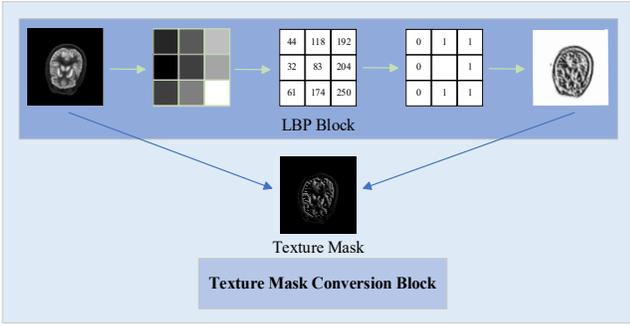

Fig. 3. The pipeline of Texture Mask Conversion Block architecture.

### E. LPEB

To facilitate accurate tracer separation in dual-tracer PET imaging, the MSP is introduced within the MS-CDT framework. Each latent space corresponds to an individual tracer, thereby enabling independent semantic and structural representation. The multi-tracer PET image and its corresponding single-tracer counterparts are embedded into a latent space with dimensions $256 \times N$, where $N$ denotes the number of tracers. When it comes to dual-tracer separation, $N = 2$. As shown in Fig. 4, it depicts the extraction process of LPEB during the training procedure and the shapes of the extracted MSP.

During the diffusion stage, compared with traditional diffusion models, the low-dimensional latent space structure of the MSP enables the MS-CDT to achieve robust estimations with fewer iteration times and a smaller model size. Traditional diffusion models incur substantial computational costs during the iterative process, thus requiring random sampling of time steps for denoising optimization. Moreover, due to the lack of joint training between the denoising process and the decoder, it implies that minor estimation errors in the denoising process can impede the full potential of the transformer. The pipeline of LPEB is illustrated in Fig. 4, mainly consists of stacked residual blocks and linear layers, along with 3x3 convolutional layers, Leaky Rectified Linear Units (LRelu), Average Pooling (Avg Pool), and Linear layers. Initially, the dual-tracer PET and the two single-tracer PETs are concatenated respectively, followed by downsampling through the pixel un-shuffling operation. Subsequently, the downsampled inputs undergo latent prior extraction to obtain the $MSP_i$, which is denoted as $L_i$. Then, the $L_i$ is employed as the dynamic modulation parameter in the multi-head transposed attention and gated feed-forward network of the transformer stage to guide the tracer PET training:

$$M' = W_l^1 L_i \odot \text{Norm}(M) + W_l^2 L_i \quad (8)$$

where $\odot$ indicates element-wise multiplication, Norm denotes layer normalization, $W_l$ represents linear layer, $M$ and $M'$ are input and output feature maps respectively.

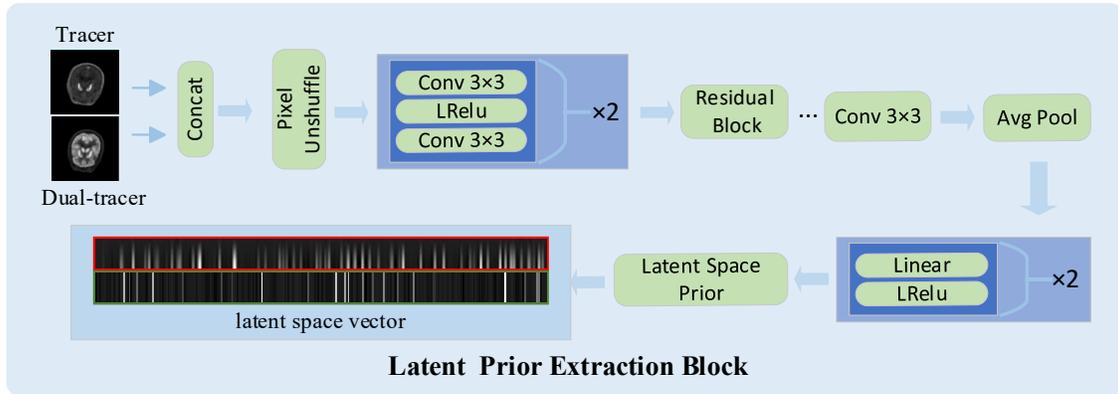

Fig. 4. The pipeline of MSP Extraction Block architecture. Conv: Convolutional layer, LRelu: Leaky Rectified Linear Unit, Avg pool: Average Pooling.

### F. Diffusion Transformer Model

The transformer model possesses a relatively strong ability to model global dependencies. When dealing with complex spatial information, it can better capture the relationships between tracers and single tracer and conduct effective image separation. The transformer model consists of two parts, namely the encoder and the decoder. In addition to the two sub-layers in each encoder layer, the decoder inserts a third sub-layer, which performs multi-head attention on the output of the encoder stack.

The encoder is composed of a multi-head attention and a gated feed-forward network. In the multi-head transposed attention, project $M'$ onto the query $Q = W_d^Q W_c^Q M'$, key $K = W_d^K W_c^K M'$, and value $V = W_d^V W_c^V M'$ matrices, where $W_c$ is the $1 \times 1$ point-wise convolution and $W_d$ is the $3 \times 3$ depth-wise convolution. Subsequently, the transposed attention map is generated through computing the dot products of the query with all keys, which is equivalent to performing the operation of aggregating global information. Further process this map using the learnable scaling parameter $\gamma$ and channel separation to generate the attention map:

$$\hat{M} = W_c \hat{V} \cdot \text{Softmax}(\hat{K} \cdot \hat{Q}/\gamma) + M \quad (9)$$

where Softmax is an activation function that converts a vector of values into a probability distribution, where each value's probability is proportional to the exponential of the input value, typically used in the output layer of a classification neural network [34].

In the gated feed-forward network, in order to aggregate local features, we utilize 1×1 Conv layers to aggregate information from different channels and 3×3 depth-wise Conv layers to aggregate information from spatially adjacent pixels. Moreover, the gating mechanism therein is employed to enhance information encoding. The characteristics of the gated feed-forward network are manifested in the following processes:

$$\hat{M} = \text{GELU}(W_d^1 W_c^1 M') \odot W_d^2 W_c^2 M' + M \quad (10)$$

where the GELU (Gaussian Error Linear Unit), which serves as an activation function. To enable the Transformer stage to effectively utilize the MSP for tracer PET separation, it employs a smooth, non-linear transformation based on the Gaussian cumulative distribution function to activate the neurons in the neural network [35], and conducts joint training on the MSP extraction block and the transformer stage.

In the diffusion stage, the strong data estimation ability of the diffusion stage is utilized to train the MSP. The MSP extraction block is employed to capture $L$, and then sampling of $L_T$ is performed through the diffusion process:

$$q(L_T | L) = \mathcal{N}(L_T; \sqrt{\bar{\alpha}_T} L, (1 - \bar{\alpha}_T)I) \quad (11)$$

$$\alpha_T = 1 - \beta_t; \bar{\alpha}_T = \Pi_{i=0}^t \alpha_i \quad (12)$$

where $T$ is the total number of iterations, $\beta_t$ indicates the predefined scale factor.

In the reverse process of MS-CDT, due to the fact that the MSP is a low-dimensional vector, it can achieve fairly good estimations with fewer iterations and a smaller model size compared to traditional diffusion models. Since traditional diffusion models lack the joint training of the denoising network and the decoder, it will result in a relatively small estimation error of the denoising network, thereby preventing the disturber from realizing its full potential. However, MS-CDT starts from the $T$-th time step (Equation (11)) and conducts all the denoising iterations (Equation (13)) to obtain $\hat{L}$, and then sends it to the Transformer stage for joint optimization.

$$\hat{L}_i^{t-1} = \frac{1}{\sqrt{\alpha_t}}(\hat{L}_i^t - \epsilon \frac{1-\alpha_t}{\sqrt{1-\bar{\alpha}_t}}) \quad (13)$$

where symbol $\epsilon$ represents the noise in $L$, and we use the LPEB and denoising network to predict noise. After that, MS-CDT eliminates variance estimation, which is conducive to accurate MSP estimation and performance improvement.

A latent representation $L$ is first extracted from the tracer PET using the LPEB during the diffusion stage reverse process. Subsequently, the denoising process is employed to estimate the noise at each time step, denoted as $\epsilon_t$. The $\epsilon_t$ is incorporated into Equation (13) to obtain $\hat{L}_i^T$, which serves as the starting point for the subsequent iterations.

After $T$ iterations, we obtain the finally estimated MSP $\hat{L}_i^T$. By jointly using $Loss$ to train the LPEB, the diffusion stage and the transformer stage, the separated data of the tracer PET can be achieved.

$$LOSS_{TM} = \|I_{Dual} - \hat{I}_{DOPA}\|_1 + \|I_{Dual} - \hat{I}_{FDG}\|_1$$
$$+ \|U_{Dual} - \hat{U}_{DOPA}\|_1 + \|U_{Dual} - \hat{U}_{FDG}\|_1 \quad (14)$$

$$LOSS_{DM} = \frac{1}{4C'} \sum_{i=1}^{4C'} |\hat{L}(i) - L(i)|, \quad (15)$$

$$LOSS = LOSS_{DM} + LOSS_{TM} \quad (16)$$

In summary, the MS-CDT has two reconstruction stages: the diffusion stage and the transformer stage. During the diffusion stage, only the MSP is trained independently, and there is no need to train the tracer PET data. In the transformer stage, the MS-CDT directly utilizes the TMC and the MSP of the tracer PET to guide the training of the transformer, with the focus on the complete PET data.

## IV. EXPERIMENTS

MS-CDT presents a flexible and computationally tractable approach for tracer separation based on the diffusion transformer model. In this section, we meticulously describe the intricate implementation details of the proposed MS-CDT, shedding light on the underlying mechanisms and techniques that drive its functionality and the datasets we utilized for evaluation. Subsequently, the reconstruction results are reported and analyzed, providing a comprehensive investigation into the performance of MS-CDT through both quantitative and qualitative evaluations. Patient data is employed for training and ablation studies.

### A. Data Specification

The experimental data consists of static 3D MAC brain data provided by the First Affiliated Hospital of Sun Yat-sen University. Data of two tracers, $^{18}F-FDG$ and $^{18}F-DOPA$, were obtained through GE Healthcare systems, with a reconstruction field of view of 700/198. In addition, the chest data for the generalization experiment was provided by the NO.2 Hospital of Anhui Medical University. Data of two tracers, $^{18}F$ and $^{68}Ga$, were acquired by scanning with the uMI510 PET/CT imaging system, and the reconstruction field of view is 500.

*Patient data:* The data from Sun Yat-sen University is a comprehensive dataset that includes 43 patients. Each patient was injected with $^{18}F-FDG$ and $^{18}F-DOPA$ tracers at different times and then scanned. The complete sampling scan time ranged from 45 seconds to 3 minutes. Since there was no dual-tracer PET data originally, the dual-tracer PET data was obtained by flexibly registering the $^{18}F-FDG$ and $^{18}F-DOPA$ tracers using ITK-snap software, and then fusing the data through 3D-Slicer software. All source images had the same spatial resolution of 256×256 pixels, and each pair of images was accurately registered. The training data included 2D slices of size $256 \times 256 \times 71$ from the horizontal sections of 38 patients, while the test data consisted of data from 5 patients, including 2 with neurological diseases and 3 healthy individuals' brain data. The data from the NO.2 Hospital of Anhui Medical University included data from 35 patients. Each patient was injected with $^{18}F$ and $^{68}Ga$ tracers at different times and scanned. The dual-tracer data was also obtained by flexible registration and then fused using 3D-Slicer software. The training data had 4,360 slice data, each with a size of $256\times256$, and the test data had 120 slice data.

### B. Model Training and Parameter Selection

In the experiment, MS-CDT is implemented in Python and PyTorch on a personal workstation with a GPU card (NVIDIA RTX-3090-24GB). This method employs a 4-level encoder-decoder structure. In the Transformer stage, the multi-head transposed attention mechanism utilized attention heads with a [1, 2, 4, 8] configuration, accompanied by respective channel numbers of [48, 96, 192, 384]. Specifically, from level 1 to level 4, the number of dynamic Transformer blocks is configured as [3, 5, 6, 6]. Additionally, the number of channels in the MSP extraction block is set to 64. In the diffusion stage, T is set to 4. The Adam optimizer is set with $\beta_1 = 0.9$ and $\beta_2 = 0.99$.

### C. Quantitative Indices

In order to assess the quality of the reconstructed data, the Peak Signal-to-Noise Ratio (PSNR), Structural Similarity Index (SSIM), Normalized Root Mean Square Error (NRMSE), Contrast Ratio (CR), and Coefficient of Variation (COV) are adopted for quantitative evaluation. PSNR is one of the metrics for measuring image quality. PSNR is defined based on MSE (Mean Squared Error). Given an original image I with a size of $m \times n$ and a noisy image $K$ obtained by adding noise to it, the MSE can be defined as:

$$MSE = \frac{1}{mn}\sum_{i=0}^{m-1}\sum_{j=0}^{n-1}[I(i,j) - K(i,j)]^2 \quad (17)$$

PSNR describes the maximum possible power related to the signal and the noise corruption power. A higher PSNR indicates better image quality. Denoting $x$ and $y$ to be the estimated reconstruction and ground-truth, PSNR is expressed as:

$$PSNR(x,y) = 20\log_{10}[Max(y)/\|x-y\|_2] \quad (18)$$

The SSIM can measure the degree of image distortion and also the similarity between two images. Unlike MSE and PSNR which measure the absolute error, SSIM is a perceptual model, that is, it is more in line with the intuitive perception of the human eye. In this paper, SSIM is used to measure the similarity between the ground truth and the corresponding images after tracer separation, and it is defined as:

$$SSIM(x,y) = \frac{(2\mu_x\mu_y+c_1)(2\sigma_{xy}+c_2)}{(\mu_x^2+\mu_y^2+c_1)(\sigma_x^2+\sigma_y^2+c_2)} \quad (19)$$

where $\mu_x$ and $\sigma_x^2$ are the average and variances of $x$. $\sigma_{xy}$ is the covariance of $x$ and $y$. $c_1$ and $c_2$ are used to maintain a stable constant. NRMSE is employed to evaluate the errors and it is defined as:

$$NRMSE(x,y) = \sqrt{\sum_{i=1}^{W}\|x_i - y_i\|_2/W}/(y_{\max} - y_{\min}) \quad (20)$$

where $W$ is the number of pixels within the reconstruction result. If NRMSE approaches to zero, the reconstructed image is closer to the reference image.

In addition to the above-mentioned metrics, contrast ratio (CR) and coefficient of variation (COV) are included in the evaluation for comprehensive assessment.

The calculation of CR involves comparing the maximum pixel value $Max_{white}$ within the lesion region of the patient to the mean pixel value $\mu_{gray}$ in the liver region:

$$CR = Max_{gray}/\mu_{white} \quad (21)$$

COV quantifies the variation in signal intensity across a selected region, like liver, of the reconstructed data. It is computed as the ratio of the standard deviation to the mean $\mu_{gray}$ of the pixel values, providing insights into the uniformity of the reconstructed image:

$$COV = \sigma_{white}/\mu_{white} \quad (22)$$

By incorporating CR and COV alongside PSNR, SSIM, and NRMSE, a more comprehensive evaluation of the reconstructed data's quality is achieved, encompassing aspects of contrast and uniformity.

### D. Experimental Comparison

To evaluate the reconstruction performance of MS-CDT, the proposed MS-CDT was compared with four states of the art models, including iVAN [36], DL-based-method [37], MPRnet [38], Pix2pix [39], and DTM [29]. We implemented several deep learning based on reconstruction methods using the same training dataset. MPRnet is a supervised model using an encoder-decoder architecture for generating clear images, while Pix2pix is a supervised GAN for denoising low - dose medical images. The involved parameters were set according to the guidelines in their original papers.

In this section, PET images were used as the input during the inference stage, with three cases of $^{18}F-FDG$ and three cases of $^{18}F-DOPA$ tracer PET brain images serving as the ground truth. The PSNR, SSIM, and MSE values of the separation results from the GE Healthcare system scanner are shown in Table 1. Table 1 presents the quantitative results of MS-CDT and various comparison methods. The best PSNR, SSIM, and NRMSE values of the separated and reconstructed images are highlighted in bold. As shown, compared with the iVAN method, the DL - based method has made significant progress in separation, and has achieved some improvements, but its results are relatively limited. Pix2pix and DTM demonstrate excellent performance in the separation and reconstruction of the two tracers compared to the previous two methods. In contrast, MPRnet performs poorly, even showing regression. In comparison, the MS-CDT achieves the best results in dual - tracer separation. In terms of separating $^{18}F-DOPA$ and $^{18}F-FDG$, the PSNR and SSIM are significantly higher than those of iVAN and the DL-based method. Additionally, although Pix2pix performs well, our separation method still has a PSNR advantage of 19.136 dB and 9.784 dB for $^{18}F-FDG$ and $^{18}F-DOPA$, respectively. Compared with the DTM method, the two tracers separated by MS-CDT have more details and less noise.

As can be seen from Table 1, MS-CDT can achieve impressive average NRMSE values of 0.5348 and 2.0554 for three cases of $^{18}F-FDG$ and $^{18}F-DOPA$ patients, respectively. The separated images contain fewer artifacts and more details, and the noise is reduced. Among them, the separation effect of the $^{18}F-FDG$ tracer is extremely excellent. Although the separation effect of the $^{18}F-DOPA$ tracer is slightly inferior to that of $^{18}F-FDG$, MS-CDT has achieved visible gains in both dual - tracer separation and noise and artifact suppression.

TABLE I
SYNTHESIS COMPARISON WITH FOUR STATE-OF-THE-ART METHODS ON BRAIN DATASET IN TERMS OF PSNR, SSIM, AND NMSE.

| Method | $^{18}F-DOPA$ | | | $^{18}F-FDG$ | | |
|---|---|---|---|---|---|---|
| | PSNR | SSIM | NRMSE | PSNR | SSIM | NRMSE |
| iVAN | 28.144 | 0.3765 | 14.6108 | 42.339 | 0.8492 | 2.9020 |
| DL-based-method | 30.987 | 0.5193 | 4.1234 | 44.397 | 0.9609 | 1.0529 |
| MPRnet | 23.824 | 0.3820 | 6.6057 | 31.050 | 0.6196 | 2.3136 |
| Pix2pix | 31.541 | 0.7278 | 6.7525 | 35.496 | 0.7752 | 1.9423 |
| DTM | 38.941 | 0.8822 | 6.0305 | 53.333 | 0.9598 | 1.2380 |
| MS-CDT | **41.325** | **0.9098** | **2.0554** | **54.632** | **0.9625** | **0.5348** |

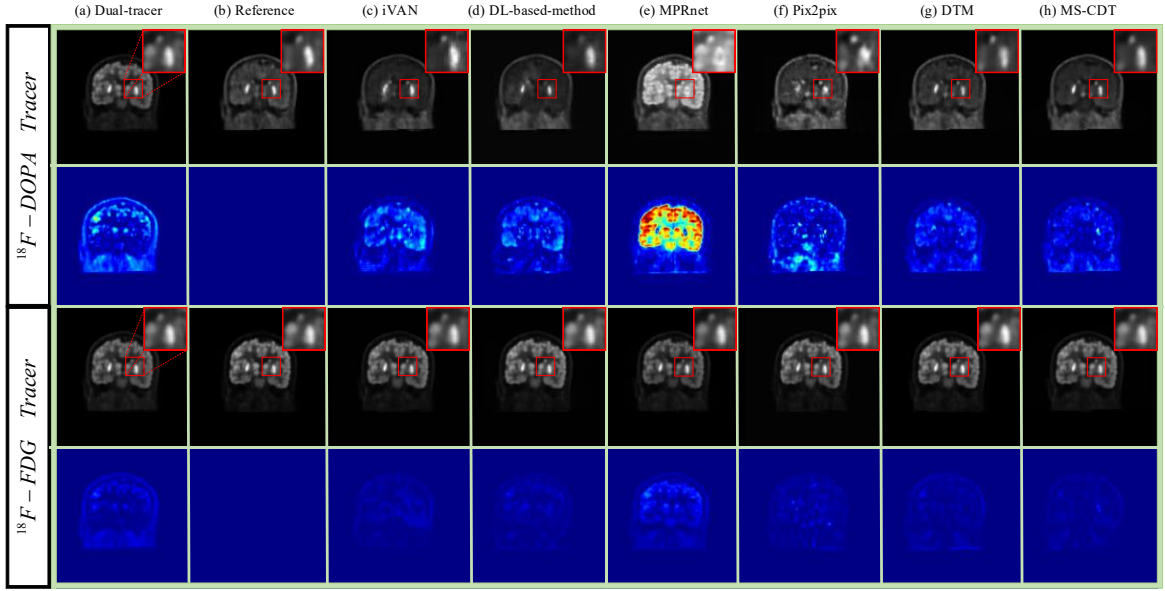

**Fig. 5.** Results of separating and reconstructing the coronal images of dual tracers of patient I, namely $^{18}F-DOPA$ and $^{18}F-FDG$ tracers, using different methods are presented. Comparisons are made among (a) the dual tracers, (b) the original PET of the tracers, (c) iVAN, (d) the DL-based method, (e) MPRnet, (f) pix2pix, (g) DTM, and (h) the images separated by MS-CDT. The detailed magnified images are described in the upper right corner of the first row. The second row shows the residuals between the reference images and the reconstructed images.

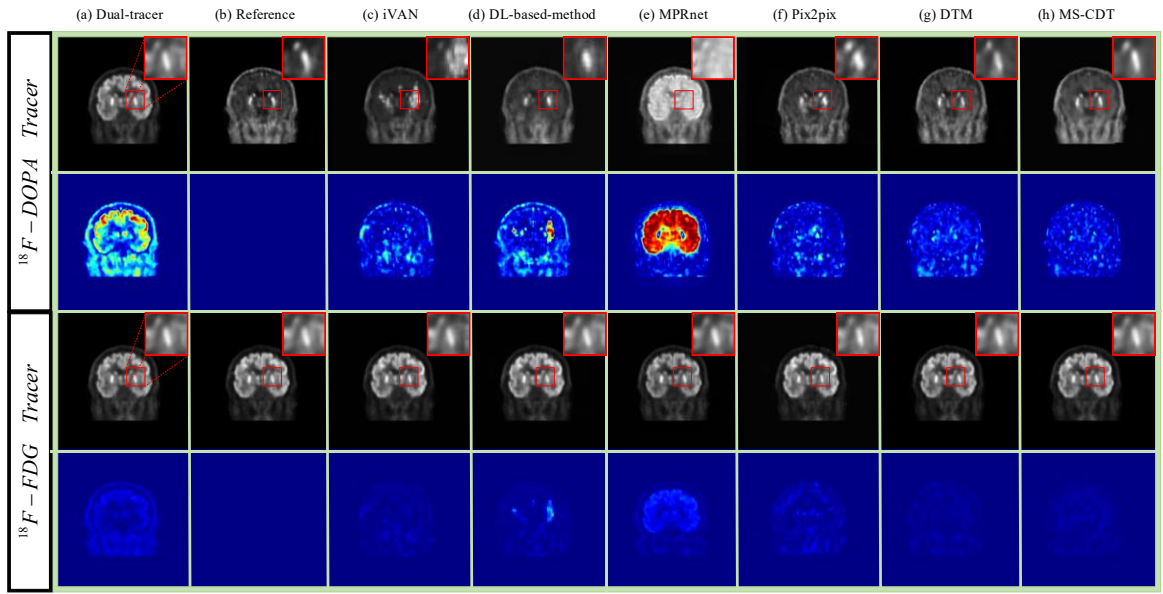

**Fig. 6.** Results of separating and reconstructing the coronal images of dual tracers of patient II, namely $^{18}F-DOPA$ and $^{18}F-FDG$ tracers, using different methods are presented. Comparisons are made among (a) the dual tracers, (b) the original PET of the tracers, (c) iVAN, (d) the DL-based method, (e) MPRnet, (f) pix2pix, (g) DTM, and (h) the images separated by MS-CDT. The detailed magnified images are described in the upper right corner of the first row. The second row shows the residuals between the reference images and the reconstructed images.

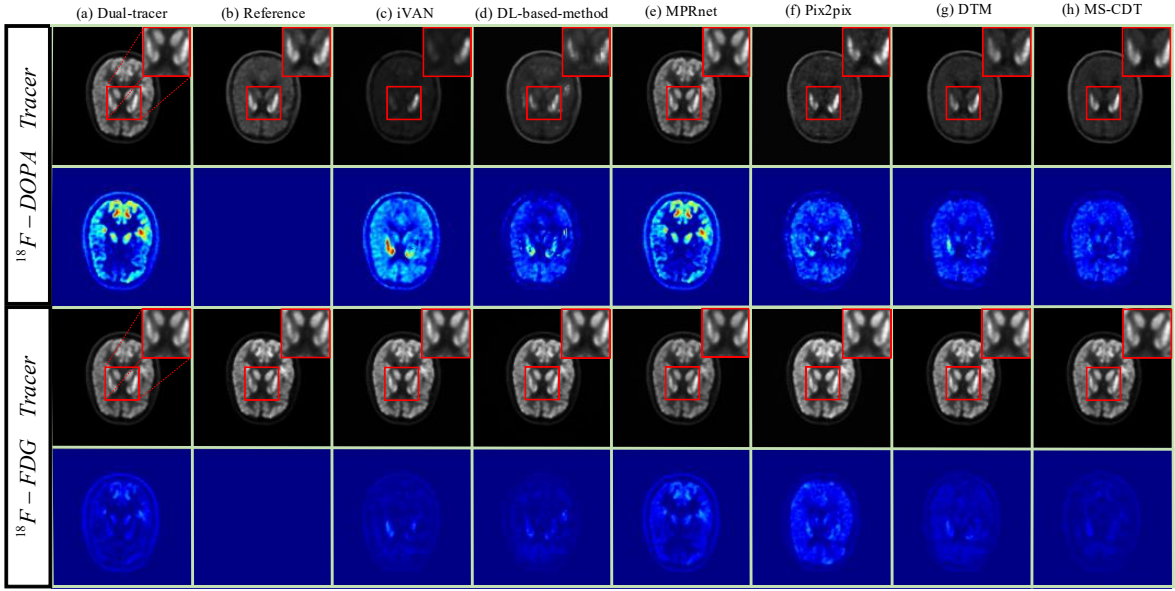

**Fig. 7.** Results of separating and reconstructing the transverse images of dual tracers of patient I, namely $^{18}F-DOPA$ and $^{18}F-FDG$ tracers, using different methods are presented. Comparisons are made among (a) the dual tracers, (b) the original PET of the tracers, (c) iVAN, (d) the DL-based method, (e) MPRnet, (f) pix2pix, (g) DTM, and (h) the images separated by MS-CDT. The detailed magnified images are described in the upper right transverse of the first row. The second row shows the residuals between the reference images and the reconstructed images.

To further illustrate the advantages of MS-CDT, analyzing the magnified rectangular regions marked with red frames allows for a deeper understanding of the performance of each method. In this extracted area, MS-CDT preserves finer details and edges, closely matching the ground truth. The error in the residual map is minimal, indicating a relatively high reconstruction accuracy. The magnified detailed images of the dual-tracer separation and the residual images in the coronal and transverse planes for the two patients are depicted in Fig. 5-8. In Fig. 5 and 6, the detailed images of the magnified transverse sections after dual-tracer separation are located in the upper-right corner of the first row, while the residual images of the transverse sections are shown in the second row. In Fig. 7 and 8, the detailed images of the magnified coronal sections after dual-tracer separation are presented in the upper-right corner of the first row, and the residual images of the coronal sections are displayed in the second row.

Although iVAN and the DL-based method successfully separated the dual tracers, their ability to reconstruct the lesion area of the $^{18}F-DOPA$ tracer was insufficient, and many details of the brain data could not be well reconstructed, introducing blurriness and inconsistency. This is particularly evident in the enlarged detailed views of the coronal and cross-sectional planes of the $^{18}F-DOPA$ tracer shown in Fig. 4-7. Moreover, the residual maps of the $^{18}F-DOPA$ tracer separated by these two methods have relatively high error values. The SAM in MPRnet incorporates dual-tracer images as additional constraints during training to correct the model output, which severely interferes with the separated images and hinders the model from learning the mapping relationship between the two tracers. Therefore, it has the worst separation performance. Pix2pix performs rather well in identifying certain detailed information of the $^{18}F-DOPA$ tracer, far exceeding iVAN, the DL-based method, and MPRnet. However, its residual map for the $^{18}F-FDG$ tracer is inferior to those of the iVAN and DL-based methods. For the DTM method, the PSNR index of the separated and reconstructed $^{18}F-FDG$ tracer is similar to that of MS-CDT, but the separation and reconstruction of the $^{18}F-DOPA$ tracer is significantly lower than that of MS-CDT. Additionally, the residual maps of both tracers are inferior to those of MS-CDT. In terms of effectiveness, the separation of both tracers by DTM is worse than that by MS-CDT, and the reconstruction of detailed information in key areas is insufficient.

In contrast, the diffusion model in our method learns the texture conditions and priors of the two tracers. The texture conditions and priors are a compressed, low-dimensional form of data representation that captures their fundamental structures, enabling independent feature disentanglement and improving the accuracy of tracer separation. Specifically, we first use the TMCB to convert the two tracers into texture features. Then, the LPEB is employed to extract the priors of the two tracers and the TMC. The diffusion model is utilized to learn the TMC and prior information. The TMCB refers to converting tracers into texture features during the reconstruction stage, guiding the model to focus on key texture features and encouraging the model to preserve fine details in the separation output. The LPEB is a technique that, during the reconstruction stage, extracts the TMC and the prior information of the two tracers. The model iteratively predicts and refines the latent space representation of the image, ensuring consistency, reducing computational complexity, and improving accuracy. During the reconstruction stage, the priors of the two tracers and the TMCs are used to assist the Transformer in directly estimating the image features, thus achieving better image separation and reconstruction quality as well as faster separation and reconstruction speed. The images reconstructed by MS-CDT are very close to the ground truth and show preserved lesions and minimal noise levels. These findings further confirm that MDTM outperforms iVAN, the

DL-based method, MPRnet, and Pix2pix in preserving details and minimizing artifacts, especially in the key areas of the images.

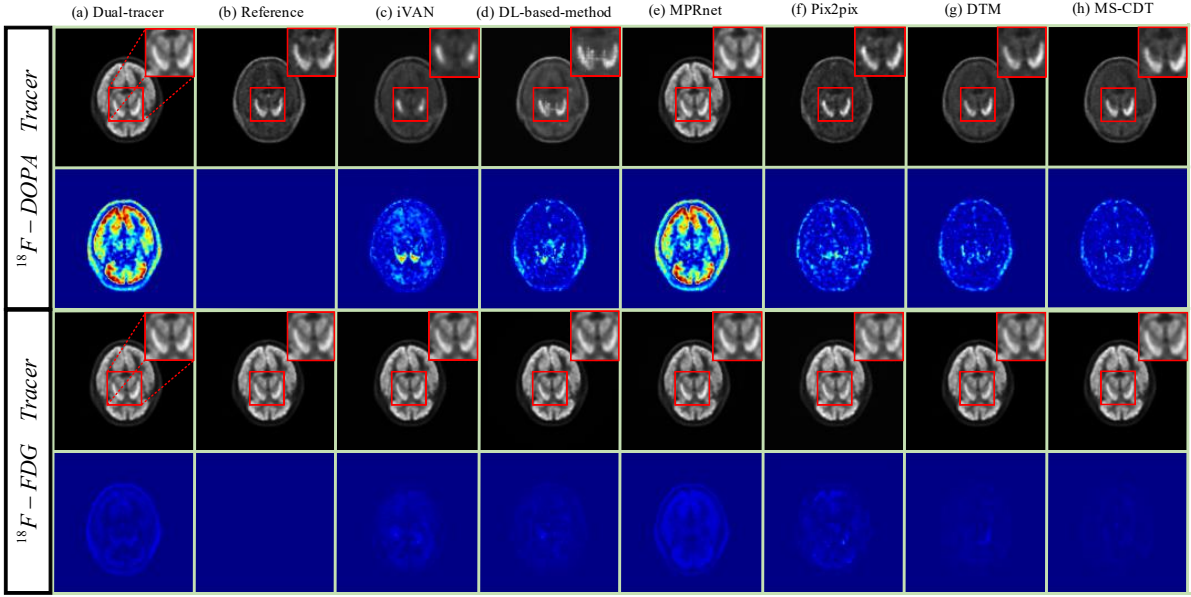

**Fig. 8.** Results of separating and reconstructing the transverse images of dual tracers of patient II, namely $^{18}F-DOPA$ and $^{18}F-FDG$ tracers, using different methods are presented. Comparisons are made among (a) the dual tracers, (b) the original PET of the tracers, (c) iVAN, (d) the DL-based method, (e) MPRnet, (f) pix2pix, (g) DTM, and (h) the images separated by MS-CDT. The detailed magnified images are described in the upper right transverse of the first row. The second row shows the residuals between the reference images and the reconstructed images.

TABLE II
RECONSTRUCT CR/COV USING THE GE HEALTHCARE SYSTEM SCANNER FOR DUAL - TRACER PET BRAIN DATA.

|  | GT | iVAN | DL-based-method | Pix2pix | DTM | **MS-CDT** |
| --- | --- | --- | --- | --- | --- | --- |
| $^{18}F-DOPA$ | 1.284/0.066 | 0.916/0.108 | 1.561/0.068 | 2.232/0.227 | 1.117/0.069 | **1.265/0.065** |
| $^{18}F-FDG$ | 6.085/0.631 | 5.328/0.634 | 5.085/0.633 | 4.008/0.639 | 5.867/0.596 | **6.144/0.589** |

Furthermore, to conduct a more comprehensive evaluation of the data quality of the two separated tracers, the CR was calculated for the 3D regions of the white and gray matter in the brains of the patients in Table 2, and the COV was evaluated for the 3D regions of the gray matter. CR aims to enhance the contrast between different tissue structures in medical images. COV is used to measure the degree of dispersion of pixel values within a certain region of an image. In Table 2, for the CR calculation, the maximum pixel value in the gray matter region of tracers $^{18}F-DOPA$ and $^{18}F-FDG$ was compared with the average pixel value in the white matter region. For the COV calculation, it was the ratio of the standard deviation of pixel values in the homogeneous area of the gray matter region of tracers $^{18}F-DOPA$ and $^{18}F-FDG$ to the average pixel value. The comparison of these two coefficients can further improve the diagnostic accuracy of medical images and assist in evaluating the uniformity of tissues.

In the CR calculation, the CR values of the DL-based-method and MPRnet deviated significantly from the CR value of the reference $^{18}F-DOPA$ data, while those of MS-CDT is very close to the reference data. By comparison, the CR value of MS-CDT was closest to that of the reference $^{18}F-DOPA$ data. For tracer $^{18}F-FDG$, the CR value of MPRnet was far from that of the reference data, while those of the DL - based - method, iVAN, and MS-CDT were close to it. Comparatively, the CR value of MS-CDT was the closest to that of the reference $^{18}F-FDG$ data, indicating that the two separated tracers were well recovered. In the COV calculation, the smaller the COV value, the lower the dispersion of image pixel values, meaning that the gray matter region is more spatially consistent. For both tracers, the COV value of MS-CDT remained at a very low level. This is particularly important in medical image processing and quality control, as a lower COV generally implies higher image quality and a lower noise level.

### E. Ablation Study

Ablation studies can provide a better understanding of the impact of each component on the performance of MS-CDT and their roles within the entire model. To further investigate the influence of certain factors on the performance of our dual-tracer separation method, we mainly discuss two factors: MSP and TMC within MS-CDT. MSP include multi-latent space priors in both horizontal and vertical directions, and TMC are used to extract the texture features of tracers.

Table IV lists the quantitative reconstruction results of each method, including PSNR, SSIM, and NRMSE. As can be seen from Fig. 7 and 8 and Table IV, MSP further improve the image quality of the model for separating dual tracers. Compared with the pure Transformer, the PSNR values increase by 4.52 dB and 2.921 dB respectively. From Fig. 7 and 8, it is evident that the model with the addition of the multi-latent space can separate and create tracers relatively well, while the pure transformer performs poorly in separating $^{18}F-DOPA$ and has difficulty in achieving effective separation. Adding the texture feature extraction module to DTM enables the model to further focus on the feature details. Compared with DTM with only MSP, the PSNR values of $^{18}F-DOPA$ and $^{18}F-FDG$ are 1.448 dB and 2.303 dB higher respectively. Overall, the results indicate that combining the multi-latent space priors and the latent priors of texture features yields the best performance in terms of both PSNR and SSIM. This further supports the effectiveness of our method in improving the quality of reconstructed and separated images by integrating these two approaches.

TABLE IV
RECONSTRUCTION PSNR/SSIM/NRMSE USING DIFFERENT COMPONENTS.

|  | Transformer | CDT | MS-CDT |
|---|---|---|---|
| $^{18}F-DOPA$ | 34.42/0.8458/3.91 | 39.84/0.8921/5.06 | 41.33/0.9098/2.06 |
| $^{18}F-FDG$ | 50.41/0.9232/1.53 | 51.46/0.9864/1.36 | 54.63/0.9625/1.30 |

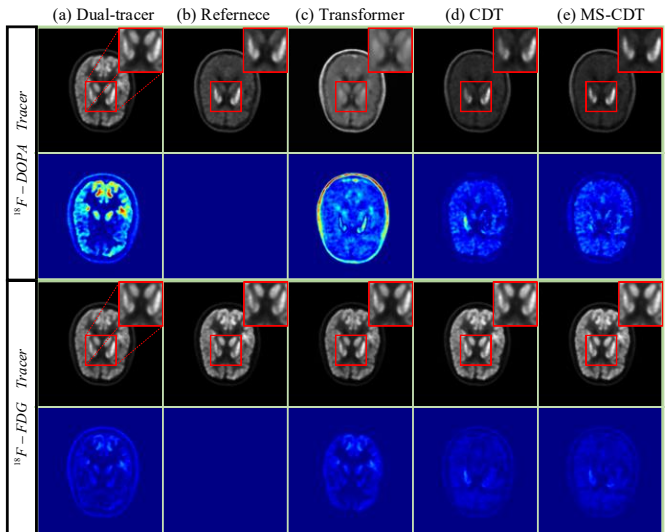

Fig. 9. Reconstruction results of cross-sectional images using different methods of patient I. (a) Dual - tracer PET, compared with (b) $^{18}F-DOPA$ and $^{18}F-FDG$ tracer PET, and (c) the separated images without MSP and TMC (d) the separated images without TMCs, and (e) the separated images by MS-CDT. The detailed magnified images are described in the upper right corner of the first row. The second row depicts the residuals between the reference images and the separated images.

### F. Generalization and Robustness Analysis

**Results of Chest Reconstruction:** To further verify the robustness of the proposed learning method, we changed the data to the dual-tracer data of $^{18}F$ and $^{68}Ga$ for the chest. There were data from a total of 35 patients, among which 4,360 images from these 35 patients were used as training data, and 120 images were used as test data. The dual-tracer data were obtained by fusing the $^{18}F$ and $^{68}Ga$ tracers using 3DSlicer. We trained and tested on the coronal plane data of the chest and evaluated its generalization performance. The calculation results obtained from the chest data are shown in Table V. The MS-CDT shows far better performance in separating the dual-tracer for the coronal plane chest data compared to other methods. In terms of the separation of the two-tracer, MS-CDT achieved the highest quantitative indicators. As can be seen from Table V, the tracer separation performance of pix2pix has significantly declined, and it is unable to effectively reconstruct the details of the tracers. The separation results of iVAN and the DL-based-method are relatively stable, but the evaluation index of NRMSE for the DL-based method fluctuates greatly. In contrast, the quantitative indicators of MPRnet have slightly improved. However, as can be seen from the separated images in Fig. 11 and 12, due to the mechanism problem of the Self-Attention Module, MPRnet has not successfully separated the tracers. It is just that the uptake of the two tracers in the lung images is relatively similar, resulting in an increase in the quantitative indicators for the unseparated case. The PSNR results of the separation and reconstruction of the two tracers by MS-CDT are at least 5.12 dB and 4.50 dB higher than those of other methods respectively. In comparison, MS-CDT has higher separation performance and stronger generalization ability, outperforming iVAN, the DL-based method, MPRnet, and Pix2pix.

As can be seen from the detailed images and residual images of the coronal plane of the separated images in Fig. 11 and Fig. 12, Pix2pix has a certain ability to suppress noise during the separation and creation processes, but its ability to separate and reconstruct the radiation area has significantly declined, leading to the worst results. MPRnet only has the ability to reconstruct and cannot effectively separate the tracers, and the results are always the dual tracer image. For iVAN and the DL-based method, the blurred internal structure and noise still exist. Compared with pix2pix, the denoising results of iVAN and the DL-based method are slightly lower, but their separation and reconstruction performance is better. In contrast, the MS-CDT method has excellent separation and reconstruction capabilities and can effectively remove noise. Compared with other competing reconstruction methods, MS-CDT stands out in reducing noise artifacts and preserving details, providing the best visual effects.

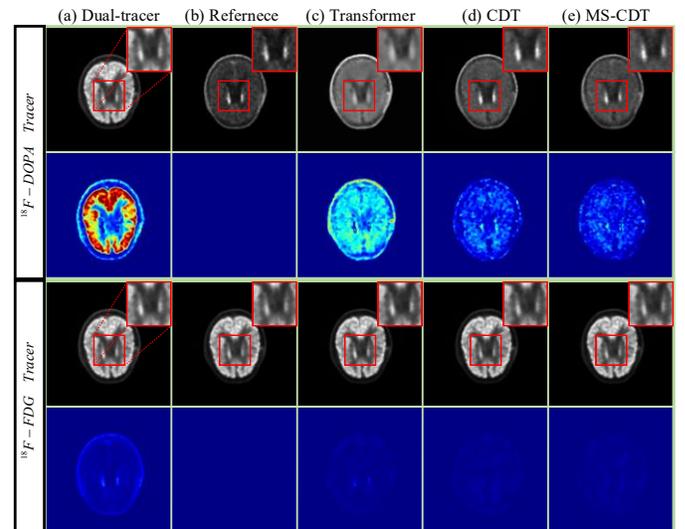

Fig. 10. Reconstruction results of cross-sectional images using different methods of patient II. (a) Dual - tracer PET, compared with (b) $^{18}F-DOPA$ and $^{18}F-FDG$ tracer PET, and (c) the separated images without MSP and TMC (d) the separated images without TMCs, and (e) the separated images by MS-CDT. The detailed magnified images are described in the upper right corner of the first row. The second row depicts the residuals between the reference images and the separated image.

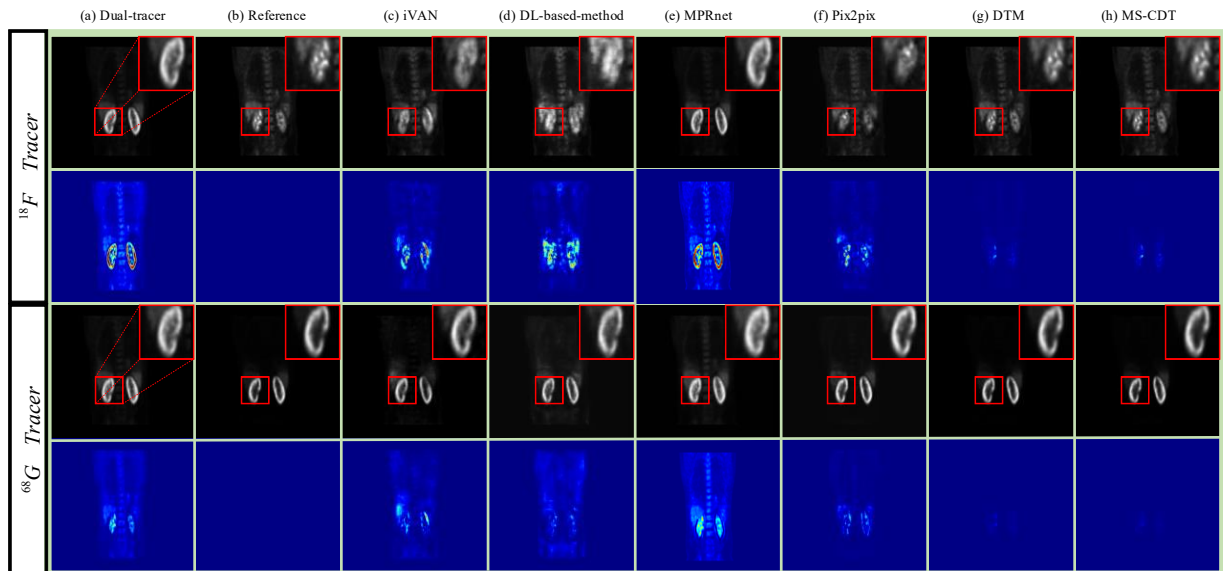

Fig. 11. Results of separating and reconstructing the transverse images of dual tracers of patient III, namely $^{18}F-DOPA$ and $^{18}F-FDG$ tracers, using different methods are presented. Comparisons are made among (a) the dual tracers, (b) the original PET of the tracers, (c) iVAN, (d) the DL-based method, (e) MPRnet, (f) pix2pix, (g) DTM, and (h) the images separated by MS-CDT. The detailed magnified images are described in the upper right transverse of the first row. The second row shows the residuals between the reference images and the reconstructed images.

TABLE V
SYNTHESIS COMPARISON WITH FOUR STATE-OF-THE-ART METHODS ON CHEST DATASET IN TERMS OF PSNR, SSIM, AND NMSE.

| Method | $^{18}F$ Tracer | | | $^{68}Ga$ Tracer | | |
|---|---|---|---|---|---|---|
| | PSNR | SSIM | NRMSE | PSNR | SSIM | NRMSE |
| iVAN | 25.10 | 0.2706 | 114.41 | 31.81 | 0.4938 | 18.14 |
| DL-based-method | 36.69 | 0.5997 | 4.74 | 35.74 | 0.5269 | 37.10 |
| MPRnet | 41.34 | 0.8498 | 28.31 | 35.31 | 0.6665 | 16.47 |
| Pix2pix | 25.32 | 0.3838 | 145.35 | 28.94 | 0.4864 | 39.63 |
| DTM | 41.39 | 0.7262 | 5.87 | 36.88 | 0.6502 | 7.57 |
| MS-CDT | **45.80** | **0.8709** | **2.73** | **39.18** | **0.6707** | **5.93** |

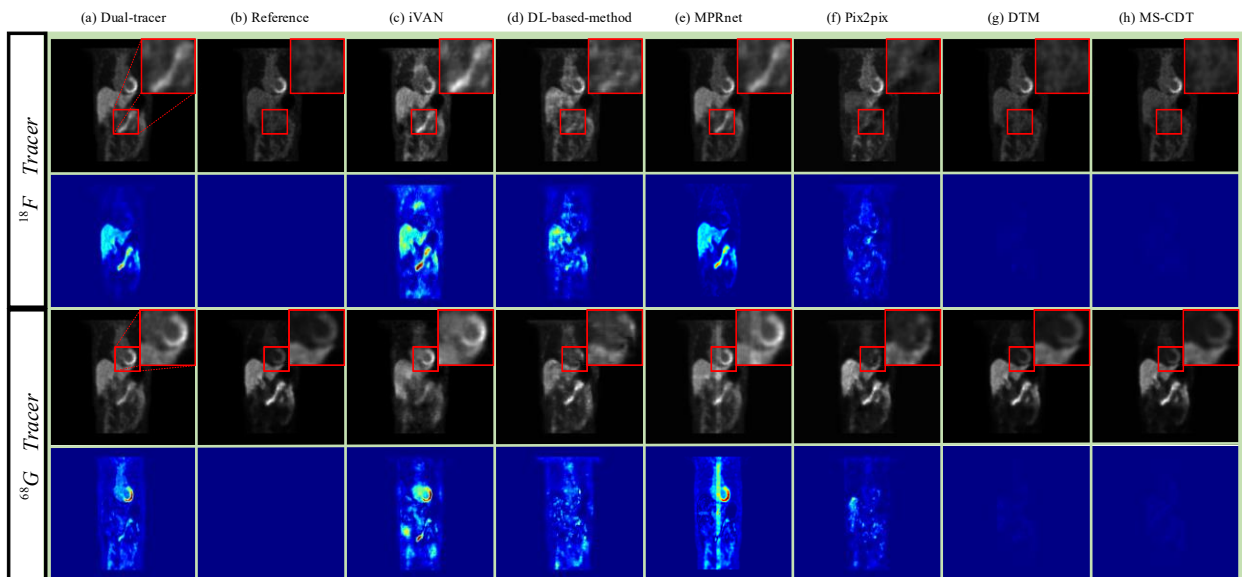

Fig. 12. Results of separating and reconstructing the transverse images of dual tracers of patient IV, namely $^{18}F-DOPA$ and $^{18}F-FDG$ tracers, using different methods are presented. Comparisons are made among (a) the dual tracers, (b) the original PET of the tracers, (c) iVAN, (d) the DL-based method, (e) MPRnet, (f) pix2pix, (g) DTM, and (h) the images separated by MS-CDT. The detailed magnified images are described in the upper right transverse of the first row. The second row shows the residuals between the reference images and the reconstructed images.

## V. DISCUSSION

The proposed MS-CDT demonstrates exceptional performance and versatility in various tracer PET separation and reconstruction tasks. This section introduces and analyzes the results of PET separation and reconstruction using texture features with different thresholds. We employed four texture feature thresholds (120, 150, 180, and 200) for tracer PET separation and reconstruction. These thresholds indicate that feature data within the ranges of (120, 255), (150, 255), (180, 255), and (200, 255) are retained, while values below these thresholds are set to zero.

Four texture feature thresholds were tested, and the results show that the texture features with a threshold of (180, 255) exhibit the best performance. Fig. 13 presents line charts of PSNR, SSIM, and NRMSE values for the four texture thresholds, while Fig. 14 and 15 display magnified striatal regions marked by red rectangles. As shown in Fig. 13-15, compared with texture features of other thresholds, the threshold of (180, 255) achieves the optimal results in PSNR, SSIM, and NRMSE among the four thresholds. Moreover, the separated tracer in the striatal region shows higher clarity and lower residuals. The objective of this experiment is to identify the optimal texture feature threshold to optimize the model's performance.

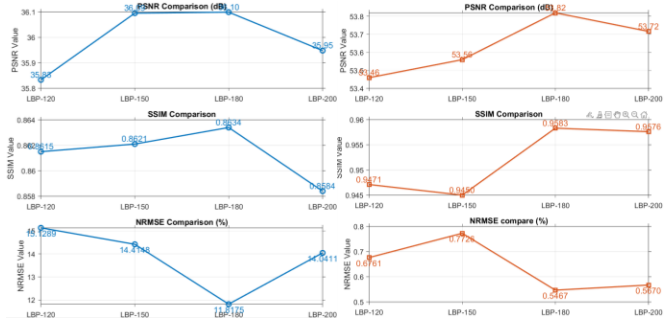

**Fig. 13.** Line charts comparing PSNR, SSIM and NRMSE of four thresholds for MS-CDT

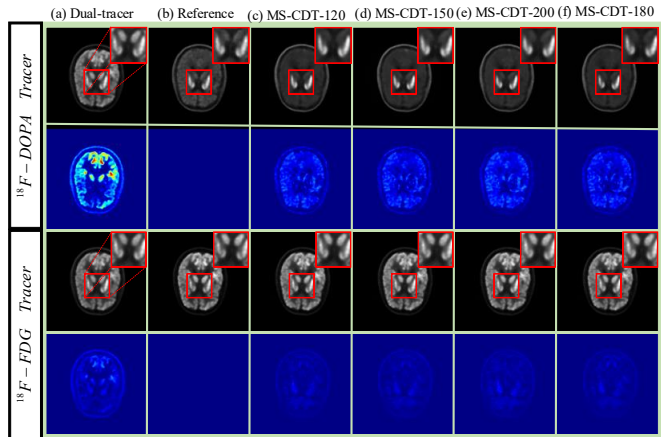

**Fig. 14.** Introduces the results of transverse images of dual tracers (i.e., Tracer A and Tracer B) for patient I reconstructed using different texture feature thresholds. A comparison is made among (a) dual tracers, (b) the original PET image of Tracer A, (c) MS-CDT_120, (d) MS-CDT_150, (e) MS-CDT_200, and (f) MS-CDT_180 images separated with different thresholds. Detailed enlarged images are described on the upper right horizontal axis of the first row. The second row shows the residuals between the reference images and the reconstructed images.

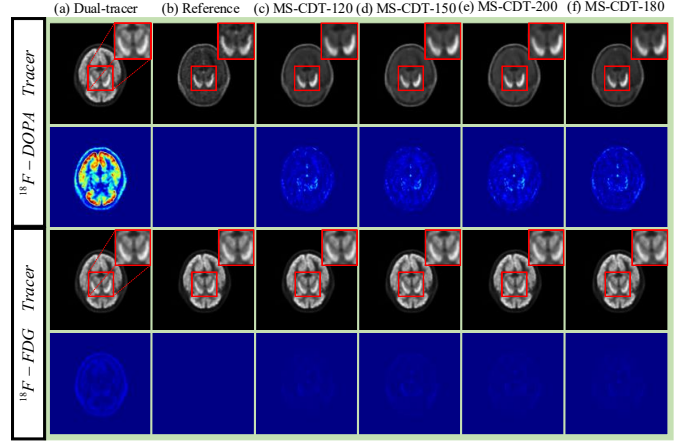

**Fig. 15.** Introduces the results of transverse images of dual tracers (i.e., Tracer A and Tracer B) for patient I reconstructed using different texture feature thresholds. A comparison is made among (a) dual tracers, (b) the original PET image of Tracer A, (c) MS-CDT-120, (d) MS-CDT-150, (e) MS-CDT-200, and (f) MS-CDT-180 images separated with different thresholds. Detailed enlarged images are described on the upper right horizontal axis of the first row. The second row shows the residuals between the reference images and the reconstructed images.

## VI. CONCLUSIONS

In clinical practice, the combined use of multiple tracers has shown great potential in enhancing the diagnosis and treatment of various diseases. However, accurately separating signals from overlapping tracer distributions remains a significant challenge. This study introduced a texture condition and multi-latent space guided diffusion transformer model for separating dual-tracer PET signals. The MS-CDT achieved independent feature disentanglement by introducing TMCs and MSP, with each TMCs and MSP corresponding to a specific tracer. This design enabled the model to more effectively decouple and extract the unique physical features of each PET tracer. By focusing on the compact TMCs and MSP instead of reconstructing the entire PET data, the diffusion process ensured more accurate separation while preserving the fundamental data features and improving computational efficiency. Furthermore, the integrated diffusion transformer model transitioned from global feature extraction to local feature recognition, achieving a balance between detail preservation and structural understanding. This study demonstrated that combining the diffusion model and the transformer model can improve the performance of multiple tracer separation and reconstruction as well as the richness of details. However, the current study has some limitations. The current method is limited to the separation of multi-tracers in the image domain. In future studies, it is essential to investigate the feasibility and effectiveness of tracer separation in the sinogram or list-mode domain, which may preserve raw physical and temporal information compared to reconstructed images. Exploring domain specific separation strategies could enable more accurate and robust disentanglement of multiple tracers.

## ACKNOWLEDGMENTS

All authors declare that they have no known conflicts of interest in terms of competing financial interests or personal relationships that could have an influence or are relevant to the work reported in this paper.